\documentstyle[twocolumn,prl,aps]{revtex}
\input epsf
\begin{document}

\draft
\twocolumn[\hsize\textwidth\columnwidth\hsize\csname@twocolumnfalse\endcsname

\title{Filler-Induced Composition Waves in Phase-Separating Polymer
        Blends } \author{Benjamin P. Lee\cite{Bucknell}, Jack
        F. Douglas\cite{corrauth}, and Sharon C. Glotzer} \address{
        Polymers Division and Center for Theoretical and Computational
        Materials Science,
        National~Institute~of~Standards~and~Technology,
        Gaithersburg,~MD~20899 } \date{ \today } \maketitle
\begin{abstract}
The influence of immobile filler particles (spheres, fibers,
platelets) on polymer blend phase separation is investigated
computationally using a generalization of the Cahn-Hilliard-Cook (CHC)
model. Simulation shows that the selective affinity of one of the
polymers for the filler surface leads to the development of
concentration waves about the filler particles at an early stage of
phase separation in near critical composition blends. These ``target''
patterns are overtaken in late stage phase separation by a
growing ``background'' spinodal pattern characteristic of blends 
without filler particles. The linearized CHC model is used to estimate
the number of composition oscillations emanating from isolated filler
particles. In far-off-critical composition blends, an ``encapsulation
layer'' grows at the surface of the filler rather than a target
pattern. The results of these simulations compare favorably with
experiments on filled phase separating blend films.
\end{abstract}


\vskip2pc]
\narrowtext

\section{Introduction}

The bulk properties of miscible fluid mixtures are characteristically
insensitive to their microscopic fluid structure near the critical
point for phase separation, where the properties are governed by large
scale fluctuations in the local fluid composition. Composition
fluctuations occur similarly for most near-critical fluid mixtures, so
that the properties of these fluids are subject to a ``universal''
description.  This accounts for the success of simple mathematical
models of critical phenomena (e.g., Ising model, $\phi^4$-field
theory) that contain the minimal physics of these fluctuation
processes. Although mixtures near their critical point are susceptible
to external perturbations, the influence of microscopic
heterogeneities tend to become ``washed out'' in the large scale fluid
properties, apart from changes in critical parameters describing the
average properties of the fluid (e.g., critical temperature and
composition, apparent critical exponents, etc.). This situation
changes, however, when the fluid mixture enters the two-phase region.
The fluid is then far from equilibrium and perturbations can grow to
have a large-scale influence on the phase separation morphology.
Perturbations in these unstable fluids can be {\it amplified} rather
than washed out at larger length scales. Inevitably, the theoretical
description of this kind of self-organization process is complicated
by various non-universal phenomena associated with the details of the
particular model or experiment under investigation \cite{langer}. The
beneficial aspect of this sensitivity of phase separation and other
pattern formation processes to perturbations is that it offers
substantial opportunities to {\it control} the morphology of the
evolving patterns and leads to a great multiplicity of
microstructures.

Many previous studies have considered the application of external
influences (flow \cite{flow} and gravitational \cite{gravity} fields,
concentration \cite{cgradient} and temperature \cite{tgradient}
gradients, chemical reactions \cite{glotzer}, crosslinking
\cite{glotzer2,qui1}, etc.) to perturb fluid phase separation, but the
investigation of geometrical perturbations is more recent.  There have
been numerous studies on the perturbation of phase separation arising
from the presence of a plane wall, which is one of the simplest
examples of a geometrical perturbation of phase separation
\cite{jones,krausch,bruder}. Measurements and simulations both show
the development of ``surface-directed'' composition waves away from
plane boundaries under the condition where one component has an
affinity for the surface. The scale of these coarsening surface waves
grows much like those of bulk phase separation patterns
\cite{jones,krausch,bruder,ball,binder,marko,krausch3,puri}. Recent
simulations \cite{karim3,kielhorn} and experiments
\cite{karim3,krausch4,boltan,ermi4,nisato} have shown that variation
of the polymer-surface interaction within the plane of the film allows
for the control of the local polymer composition in blends phase
separating on these patterned substrates (``pattern-directed phase
separation'' \cite{karim3}).  Measurements have also indicated that
the polymer-air boundary of phase-separating blend films on patterned
substrates can be strongly perturbed by phase separation within the
film \cite{nisato,karim,sung} and thermal fluctuations of the
polymer-air boundary can also strongly influence the structure of thin
polymer films \cite{xie}.

In the present paper we focus on the consequences of having
geometrical heterogeneities of finite extent in a phase separating
blend.  The Cahn-Hilliard-Cook (CHC) theory \cite{cahn} for phase
separation is adapted to describe phase separation of a blend with
spherical, cylindrical (fiber), and plate-like shaped filler
particles. The extended dimensions of the fiber and platelet filler
particles are taken to be much larger than the scale of the phase
separation process. A variable polymer-surface interaction is
incorporated into the filler model in a fashion similar to previous
treatments of plane surfaces \cite{ball,binder,marko}. 

The paper is organized as follows.  In section II we briefly summarize
the CHC model to introduce notation, to define the relation between
model parameters and those of polymer blends theory, and to explain
modifications required for incorporating immobile filler particles
into the CHC simulations of phase separation. Section III summarizes the
results of simulations for representative situations.  Key phenomena
are identified: 1) Target composition patterns form in near critical
composition blends. 2) Target patterns are a transient phenomenon. 3)
The scale of the target patterns depends on quench depth and molecular
weight. 4) Qualitative changes in the filler-induced composition
patterns occur when the surface interaction is neutral and when the
blend composition is off-critical. 5) Multiple filler particles induce
composition waves exhibiting complex interference patterns.  Section IV
provides a simple analytic estimate of the scale of the target pattern
based on the linearized CHC theory, and these results are tested
against simulations for circular filler particles.  In section V
simulation results are compared to experiments on ultrathin polymer
films having silica bead filler particles immobilized by the solid
substrate on which the films were cast.  Atomic force microscopy
measurements on the filled blend films are compared to the analytical
predictions. Simulations of phase separation in off-critical blend
films are briefly compared to analogous experiments.  Measurements on
crosslinked blend films are also considered.  The final section (VI)
discusses generalizations of the present study to manipulate the
structure of phase-separating blends.

\section{The Model} 

\subsection{CHC Equation}

We present a brief discussion of the CHC model
\cite{cahn,glotzerreview,otherreviews} to introduce notation and to explain
the modifications required for incorporating filler particles. The
modeling of the phase separation dynamics is based on gradient flow
of a conserved order parameter $\phi({\bf r},t)$,
\begin{equation}\label{EQN:Langevin}
        {\partial\over\partial t}\phi({\bf r,t}) = M \nabla^2 
        {\delta F[\phi] \over \delta\phi({\bf r})}
        + \zeta({\bf r},t),
\label{chc}
\end{equation}
with $\phi({\bf r},t)$ equal to the local volume fraction of one of the
blend components. Incompressibility of the mixture is assumed so that
the local volume fraction of the second component is $1-\phi$.  The mobility
$M$ is assumed to be spatially uniform and independent of
concentration and the free energy functional $F[\phi]$ has the general form
\begin{equation}\label{EQN:functional}
        F[\phi({\bf r})] =\int {d{\bf r}\over v}
        \left[ {\textstyle{1\over 2}} k_BT \kappa(\phi) (\nabla \phi)^2 
        + f(\phi) - \mu_{eq} \phi \right],
\end{equation}
where $f(\phi)$ is the bulk Helmholtz free energy per lattice site,
$v$ is the volume per lattice site, $k_B$ is Boltzmann's constant, $T$
is temperature, and $\kappa(\phi)$ is a measure of the energy required
to create a gradient in concentration.  Higher order gradient terms
are neglected.  The chemical potential is given by $\mu_{eq} =
\partial f / \partial \phi |_{\phi_{eq}}$, which ensures that
$\phi({\bf r}) = \phi_{eq}$ is the solution of $\delta
F[\phi]/\delta\phi({\bf r})=0$.  Finally, thermal fluctuations
necessary to ensure a Boltzmann distribution of $\phi({\bf r})$ in
equilibrium are included via the Gaussian random variable $\zeta$.
The average of $\zeta$ vanishes, and $\zeta$ obeys the relation 
\begin{equation}\label{EQN:noise}
        \langle \zeta({\bf r},t) \zeta({\bf r}',t') \rangle =
        - 2 M k_B T \nabla^2 \delta({\bf r}-{\bf r}') \delta(t-t').
\end{equation}

For temperatures near the critical temperature $T_c$ the free energy
can be expanded in powers of the composition fluctuation $\psi({\bf
r})=\phi({\bf r})-\phi_c$, giving the Ginzburg-Landau (GL) functional
\begin{equation}\label{EQN:GLfunctional}
        F[\psi({\bf r})] = k_BT \int {d{\bf r}\over v} 
        [ {\textstyle{1\over 2}}\kappa_c
        (\nabla\psi)^2 + {\textstyle{1\over 2}}c \psi^2 +
        {\textstyle{1\over 4}} u \psi^4 + \dots ]
\label{gl}
\end{equation}
where $\kappa_c\equiv \kappa(\phi_c)$.  Neglected terms are either
higher order in $\psi$ or in $\sqrt c \propto \sqrt{T-T_c}$, which is
small near the critical point.  Eq.~(\ref{chc}), in combination with 
(\ref{gl}), defines the well-known Model B \cite{hh},
\begin{equation}\label{EQN:CHC}
        {\partial \over \partial t}\psi({\bf r},t) = 
        - M \left(k_BT\over v\right)\nabla^2 
        ( \kappa_c \nabla^2 \psi - c\psi - u \psi^3 ) 
        + \zeta({\bf r},t),
\label{modelB}
\end{equation}
Eq.~(\ref{EQN:CHC}) is used to study the dynamics following a
quench to the two-phase region $T < T_c$, where $c < 0$.  In that
case, the CHC equation may be rescaled into the dimensionless form
\cite{grant}
\begin{equation}\label{EQN:CHCdimless}
        {\partial \over \partial t} \psi({\bf r},t) = - \nabla^2
        (\nabla^2 \psi + \psi - \psi^3) 
        + \epsilon^{1/2}\eta({\bf r},t)
\end{equation}
by making the substitutions 
\begin{eqnarray}
        {\bf r}         & \to &  (|c|/\kappa_c)^{1/2} \, {\bf r} \nonumber \\
        t               & \to &  (M k_BT c^2/v\kappa_c) \> t \label{EQN:t}\\
        \psi            & \to &  (u/|c|)^{1/2} \, \psi \nonumber
\end{eqnarray}
Note that this amounts to rescaling space by $\sqrt 2 \xi_-$, where
$\xi_-$ is the thermal correlation length in the two-phase region, and
time by $\tau= D_{coll}/(2\xi_-^2)$, with $D_{coll}$ the collective
diffusion coefficient.  Here the noise term $\eta({\bf r},t)$
satisfies $\langle \eta({\bf r},t) \rangle = 0$ and
\begin{equation}\label{EQN:dimlessnoise}
        \langle \eta({\bf r},t) \eta({\bf r}',t') \rangle =
        - \nabla^2 \delta( {\bf r} - {\bf r}') \delta( t - t').
\label{noiseeqn}
\end{equation}
The only parameters left to specify the dynamics are the (conserved)
average concentration $\psi_0 \equiv \langle\phi\rangle-\phi_c$ and
the dimensionless noise strength parameter $\epsilon$,
\begin{equation}
        \epsilon = 2 u / (\kappa_c^{d/2}  |c|^{(4-d)/2}).
\end{equation}
Roughly speaking, the reciprocal of $\epsilon$ is a measure of
quench depth.  The parameter $\epsilon$ also arises in discussions
of the width of the critical region, and the connection between
thermal noise strength and the Ginzburg criterion was first noted by
Binder \cite{binder2}.

\subsection{Polymer Blends}

For polymer blends we take the Flory-Huggins (FH) form of the 
Helmholtz free energy per lattice site,
\begin{equation}\label{EQN:FloryHuggins}
        {f^{FH}(\phi)\over k_BT} 
        = {\phi\over N_A} \ln \left( \phi\over N_A \right)
        + {1 - \phi \over N_B} \ln \left( 1 - \phi \over N_B \right)
        + \chi \phi (1 - \phi).
\end{equation}
Here $\chi$ represents the monomer-monomer interaction energy, $N_i$
is the polymerization index of component $i$, and $\phi$ is the volume
fraction of component A.  For the coefficient of the gradient term we
use de Gennes' random-phase approximation (RPA) result (neglecting the enthalpic contribution \cite{glotzerreview}), 
\begin{equation}
        \kappa(\phi) = {1\over 18} \left[ {\sigma_A^2 \over \phi}
        + {\sigma_B^2 \over 1 - \phi} \right],
\end{equation}
where $\sigma_A$ and $\sigma_B$ 
are monomer sizes of the $A$ and $B$
blend components, given in terms of the radius of gyration of the $i$th
component $R_{g,i}$ by $\sigma_i^2 = R_{g,i}^2/N_i$.

FH theory exhibits a critical point at 
\begin{eqnarray}
        \phi_c  & = &           N_B^{1/2}/(N_A^{1/2}+N_B^{1/2}) \nonumber \\
        \chi_c  & \equiv &      \chi(\phi_c,T_c) =
                          \bigl[N_A^{1/2}+N_B^{1/2} \bigr]^2 / (2 N_A N_B). 
\end{eqnarray}
Consequently, the coefficients of the GL functional are defined as
\cite{chempot}
\begin{eqnarray}
        c               & = &   2 \chi_c (1 - \chi/\chi_c) \nonumber \\
        u               & = &   {4\over 3} \chi_c^2 \sqrt{N_A N_B} \nonumber \\
        \kappa_c        & = &   {1\over 18} \left[\sigma_A^2 \left(1+\sqrt{N_A/N_B}\right) +
                                \sigma_B^2 \left(1+\sqrt{N_B/N_A}\right) \right].
\end{eqnarray}
\noindent The phase separation dynamics of polymer blends can then be described
by the dimensionless CHC equation with $\epsilon$ determined by
molecular parameters,
\begin{equation}\label{EQN:epsilon}
        \epsilon = {36\sqrt 2 f(x) \over \gamma^3
        (\chi/\chi_c - 1)^{1/2} \, N^{1/2} },  
\end{equation}
where $N_A = N$ and $x=N_B/N$, $f(x) = (1 + \sqrt{x})^3 / 8 x$, and
$\gamma$ is the ratio of monomer size to lattice size,
\begin{equation}
\gamma={\left[\sigma_A^2(1+1/\sqrt x)+\sigma_B^2(1+\sqrt x)\right]^{1/2}
        \over 2 v^{1/3}}. 
\end{equation}
$\gamma$ simplifies to $\gamma=\sigma_{A,B}/v^{1/3}$ for symmetric blends.
Thus we see that deep quenches ($\chi \gg \chi_c$) and high molecular 
weight polymers effectively reduce the thermal fluctuations in the rescaled
dynamical equations.

\subsection{Surface Energetics}

In the presence of a surface we add a local surface interaction energy
to be integrated over the boundary,
\begin{equation}
        F_s[\psi] 
        = \int_S d^{d-1}x \, [h \psi + {\textstyle{1\over 2}}g \psi^2 
        + \dots].
\end{equation}
The coupling constant $h$ in the leading term plays the role of a
surface field which breaks the symmetry between the two phases, i.e.,
attracts one of the components to the filler surface.  The coupling
constant $g$ in the second term is neutral regarding the phases, and
results from the modification of the interaction energy due to the
missing neighbors near the surface \cite{missing} and chain
connectivity \cite{chain}.  For studies of surface critical phenomena
\cite{diehl,dietrich} and surface dynamics \cite{puri3,diehl2} one
typically keeps only these terms as a minimal model of phase
separation with boundaries.

There has been considerable attention given to the subject of the 
appropriate dynamical equations for the surface boundary conditions 
\cite{puri3}.  We follow most authors and impose zero flux at the
boundary, $\hat n\cdot {\bf j}_\psi=0$, which gives
\begin{equation}\label{EQN:conservation}
        \hat n\cdot\nabla (\nabla^2\psi + \psi - \psi^3) = 0.
\end{equation}
For the second condition we impose that of local equilibrium at the surface,
namely,
\begin{equation}\label{EQN:bcii}
        \hat n\cdot \nabla \psi = h + g\psi.
\end{equation}
While more sophisticated treatments are available \cite{puri}, they
lead to dynamics which rapidly relax to equilibrium and
satisfy the above condition.  The details of how we implement
(\ref{EQN:conservation}) and (\ref{EQN:bcii}) in simulations with a
curved interface are presented in Appendix A.

\subsection{Simulation Details}

The equation of motion is solved using a standard central finite
difference scheme for the spatial derivatives, and a first-order Euler
integration of the time step \cite{press}.  In all the simulations,
the lattice spacing is taken between $0.7$ and $1.0$ in dimensionless
units (sufficiently smaller than any relevant physical length scales)
and the time step is taken sufficiently small to avoid numerical
instability. In the present paper, all simulations are performed in
$d=2$ on lattices up to size 128$^2$, depending on the choice of mesh
size. We note that the CHC equation is known to exhibit quantitatively
similar pattern formation and coarsening kinetics in two and three
dimensions. Further technical details about this type of simulation
can be found in Ref.~\cite{glotzerreview}.

\section{Illustrative Simulations} 

\begin{figure}
\hbox to\hsize{\epsfxsize=0.8\hsize\hfil\epsfbox{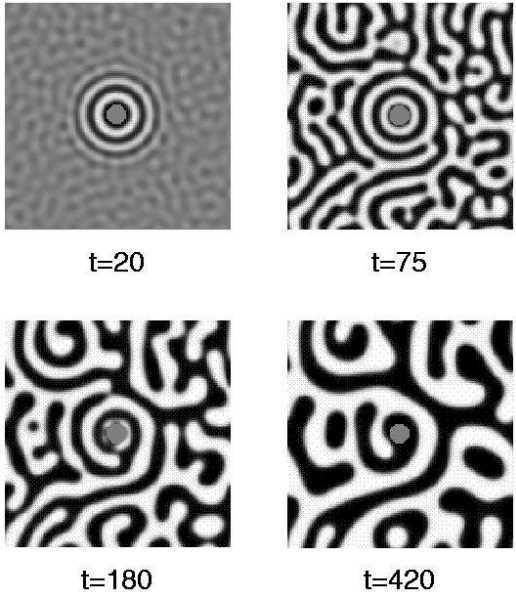}\hfil}
\smallskip
\caption{CHC simulation of the influence of filler particles on
polymer blend phase separation in a critical composition
blend. Calculations are performed in $d=2$ with $\epsilon=10^{-5}$,
$g=1.0$ and $h=1.0$.  In the reduced units of CHC theory [cf.
Eq.~(\protect\ref{EQN:t})], $R=5.6$ (or in explicit units $R\simeq
3.96\xi^-$).  The target phase separation patterns are well developed
at early times, but fragment as the ``background'' spinodal phase
separation pattern coarsens to a scale larger than the filler particle
(central gray region in figure).}
\label{FIG:transient}
\end{figure}

In Fig.~\ref{FIG:transient} we show the influence of an isolated
circular filler particle on the development of the phase separation
pattern of a blend film having a critical composition. The thermal
noise is small in these two-dimensional simulations
($\epsilon=10^{-5}$), corresponding to high molecular weight and/or a
deep quench (low and high temperatures relative to the critical
temperature for upper and lower critical solution type phase diagrams,
respectively).  At an early stage of the phase separation process
the filler particle creates a spherical composition wave disturbance
that propagates a few ``rings'' into the phase separating medium in
which the filler particle is embedded.  The target rings initially
have the size of the maximally unstable ``spinodal wavelength''
$\lambda_0$ obtained from the linearized theory \cite{cahn}.  As the
characteristic scale of the bulk phase separation pattern coarsens to
the size of the filler particle, the outer rings of the ``target''
pattern become disconnected and increasingly become absorbed into the
background spinodal pattern. The perturbing influence of the particle
becomes weak at a late stage of the phase separation where the scale
of the background phase separation pattern exceeds the filler particle
size.  The finite extent of the filler thus limits the development of
the composition waves to a transient regime.

\begin{figure}
\hbox to\hsize{\epsfxsize=0.8\hsize\hfil\epsfbox{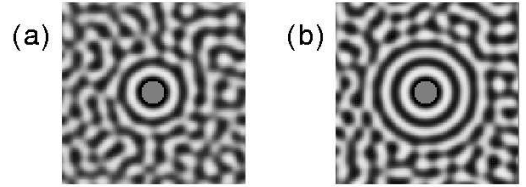}\hfil}
\vspace{5mm}
\caption{Composition waves resulting from different values of thermal noise:
 (a) $\epsilon=10^{-3}$ and time $t=21$ in dimensionless units; (b)
$\epsilon=10^{-5}$ and time $t=33$.  The surface interaction 
parameters are $h=1.0$ and $g=0$ for both, and the filler radius is $R=5.6$.}
\label{FIG:noise}
\end{figure}

The formation of target patterns can be described as composition waves
propagating into the bulk, unstable region until they are overwhelmed
by the developing background spinodal decomposition pattern.  The rate
of onset of spinodal decomposition is controlled by the strength of
thermal fluctuations, hence the noise parameter $\epsilon$ plays a
crucial role in determining the radial extent of the composition
waves.  Fig.~\ref{FIG:noise} shows two systems with equal surface
interaction, but with varying noise strengths: (a) $\epsilon=10^{-3}$
and (b) $\epsilon=10^{-5}$.  We see that the spatial extent of the
target pattern is larger for smaller $\epsilon$.  Consequently, we
expect deeply quenched and/or high molecular weight polymer blends to
be favorable systems for observing filler induced composition waves
because of the relatively low thermal noise level typical of these
systems, and the relatively high viscosity of these fluids which slows
the dynamics and makes measurements of intermediate stage patterns
possible (e.g. via atomic force microscopy as discussed in section V).
In section \ref{SEC:estimate}, we use the linearized CHC equation to
estimate the extent of the composition wave.

\begin{figure}
\hbox to\hsize{\epsfxsize=0.8\hsize\hfil\epsfbox{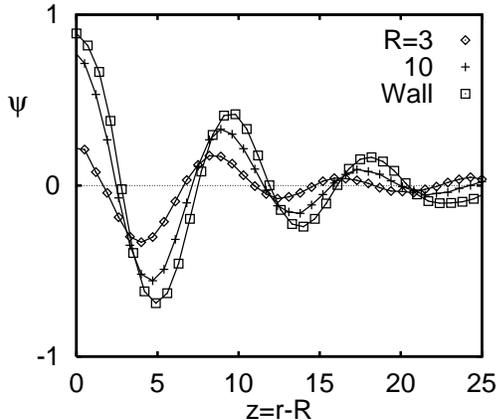}\hfil}
\smallskip
\caption{Typical angular-averaged radial composition profiles
$\psi(z)$ for $R=3, 10$ and $\infty$ in an intermediate stage of phase
separation. Parameters used are $h=0.1$, $g=0$, $\epsilon=10^{-2}$,
and the time is $t=25$. Averages were taken over ten independent
configurations.}
\label{FIG:proR}
\end{figure}

We observe that the composition waves disappear when the particle
radius $R$ becomes vanishingly small, and the persistent waves
developing from planar surfaces are recovered for very large spherical
particles.  We then examine particles of sizes intermediate between
these extreme limits and during the intermediate phase separation
period in which the target patterns are well developed.
Fig.~\ref{FIG:proR} shows the angular-averaged composition profile
$\psi(z)$, with $z\equiv r-R$ the radial distance from the surface of
the filler particle.  We observe that the amplitude of the local
composition fluctuations become more developed and more sharply
defined with increasing filler size, $R$.  In comparison to the planar
surface ($R\to\infty$), the composition wave profile for $R=10$ is
only slightly reduced in amplitude, whereas for $R=3$ the amplitude is
reduced to half that of the wall case.  We also remark that the radial
extent of the target pattern is similar for all particle sizes.

We next examine the influence of the surface interaction on
the formation of filler-induced target patterns.  The impact of the
symmetry breaking perturbation of the filler particle on the phase
separation may be tuned through the surface interaction parameters $g$
and $h$. We focus our attention on $h$ since it has a predominant
effect on the resulting pattern formation.

\begin{figure}
\hbox to\hsize{\epsfxsize=1.0\hsize\hfil\epsfbox{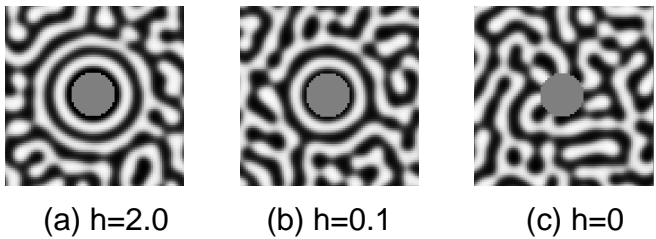}\hfil}
\vspace{5mm}
\caption{Influence of surface interaction on filler-induced pattern
formation. The figures show intermediate stage phase separation ($t
=50$) with thermal noise $\epsilon=10^{-4}$ and radius $R=9.6$.
Filler-induced target pattern for a selective polymer interaction (a)
$h=2.0$ and (b) $h=0.1$, and (c) a non-selective polymer interaction,
$h=0$.  Note the tendency for the domains to align 
perpendicularly to the neutral filler interface.}
\label{FIG:hfig}
\end{figure}

Fig.~\ref{FIG:hfig} shows CHC simulations of blend phase separation
for the case where one component strongly prefers the filler, weakly
prefers the filler, and has no preference for the filler ($h=0$). The
quench depth parameter and computation times are identical in these
images. We see that the target patterns do not form in the case of
filler particles with a (``neutral'') non-selective interaction, but
rather there is a tendency for the patterns to align {\it
perpendicularly\/} to the interface. This type of compositional
alignment, which we find is even more pronounced in the case of a
planar surface, also occurs in block copolymer fluids \cite{block}.
The perturbing influence of the boundary interaction saturates with an
increase of $h$, as can be seen by comparison of the $h=2.0$ and
$h=0.1$ cases.  Below we demonstrate that the extent of the target
pattern depends logarithmically on $h$.

Target waves are a variety of spinodal pattern with a symmetry set by the
shape of the filler particle boundary. The introduction of surface
patterns on a solid substrate can similarly break the symmetry of the
phase separation process and can be used to impart a particular
``shape'' to the spinodal pattern \cite{karim3,boltan}.  A previous
CHC simulation by us considered this ``patterned-directed'' phase
separation in near critical composition blend films \cite{karim3}.

The blend composition also can have a large influence on the character
of the filler-induced phase separation structures in blends,
particularly when sufficiently far off critical to suppress the
spinodal instability.  In this case we find a layer of composition enrichment
(``encapsulation layer'') forms about the filler particle, but there
are no target patterns \cite{benderley}. This encapsulation layer
grows in time, but appears to grow slower than $t^{1/3}$.  A
non-selective interaction ($h=0$) leads to the absence of
encapsulation by the minority phase, and minority phase nucleation
occurs largely unaffected by the presence of the filler.  We thus find
that the development of target patterns requires the conditions of
ordinary spinodal pattern formation (i.e., ``near'' critical
composition) and the existence of a heterogeneity to initiate the wave
disturbance.

\begin{figure}
\hbox to\hsize{\epsfxsize=0.7\hsize\hfil\epsfbox{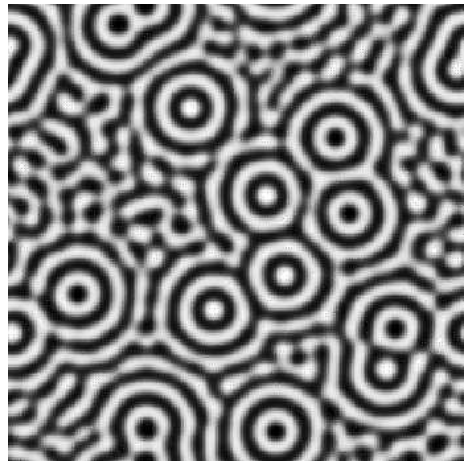}\hfil}
\vspace{5mm}
\caption{Illustrative CHC simulation of blend phase separation with
many filler particles.  Filler particles preferential to each phase
are included (the two filler types have a dark and light filler core).
Simulation parameters are $t=80$, $h=\pm 1.0$, $R=2.0$, and
$\epsilon=10^{-6}$.  Note the interference pattern between these
composition waves.}
\label{FIG:multiple}
\end{figure}

A representative example of the interference between filler-induced
rings at a non-vanishing filler concentration is shown in
Fig.~\ref{FIG:multiple}. Filler particles can each have an affinity
for the different blend components so that the enriching phase
(``charge'') can vary near the surface of the filler particle at the
core of the target waves. (Sackmann has noted that composition
enrichment patterns occur about membrane proteins in lipid mixtures
comprising living cell membranes, and these patterns mediate protein
interactions, leading to attractive or repulsive interactions
depending on the ``charge''\cite{sackmann}.) In the low noise
limit it should be possible to obtain novel wave patterns like those
found in reaction diffusion models with regularly spaced sources for
wave propagation \cite{steinbock}, but we do not pursue this here.  We
do mention that the use of filler particles responsive to external
fields could allow the manipulation of the large scale phase
separation pattern if the external fields are used to align the filler
particles.

\section{Estimation of the Spatial Extent of the Target Patterns}

\label{SEC:estimate}

In this section we derive an approximate analytical expression for the
spatial extent of the composition wave pattern.  Our method is based
on the observation above that the composition wave propagates until it
is overwhelmed by the growth of the bulk spinodal decomposition
background pattern.  In the context of surface-directed spinodal
decomposition, a qualitative explanation of this type of phenomena  was
proposed by Ball and Essery \cite{ball}, who argued that the early
time dynamics can be adequately described by the linearized CHC
equation, in which the composition wave and the bulk spinodal
decomposition add linearly.  Both processes continue independently
until a local non-linear threshold value of $|\psi|\approx \psi_t$ is
reached, which then relaxes toward the equilibrium value $\psi=\pm 1$.

We test this conjecture numerically in the case of the filler
inclusions.  First, we simulate the CHC equation in bulk, with no
filler particle, and determine the time $t_0$ at which the
root-mean-square concentration $\langle\psi^2(t)\rangle^{1/2}$ reaches
the threshold value $\psi_t=0.15$ (our reason for this choice is given below).
Next, we simulate the filled blend
{\it in the absence of noise\/} solving for the pattern at time $t_0$,
and then estimate the radius $r_0$ at which the composition wave
(envelope) exceeds $\psi_t$.  Finally, we perform the simulation with
{\it both\/} the filler particle and the thermal noise, and find the
pattern to be well-characterized by the size $r_0$ for a range of
noise and surface interaction parameters.

Based on these ideas, we next develop an analytic estimate of the spatial extent of
the pattern using the linearized CHC theory
\cite{cahn}.  At early times the order parameter does not deviate
significantly from zero, and one linearizes (\ref{EQN:CHCdimless})
to obtain
\begin{equation}\label{EQN:CHClin}
        \partial_t\psi = -\nabla^2(k_0^2 + \nabla^2)\psi + \epsilon^{1/2}\eta
\label{linearpsi}
\end{equation}
where $k_0^2 = 1 - 3 \psi_0^2$.  We apply this equation first to the 
determination of the root-mean-square order parameter 
$\psi_{\rm rms}(t)\equiv\langle\psi(t)^2\rangle^{1/2}$, which 
characterizes the rate of growth of the concentration fluctuations at very 
early times following a quench to the two-phase region.
Fourier
transformation gives
\begin{equation}
        \partial_t\tilde\psi({\bf k},t) = k^2(k_0^2 - k^2)\tilde\psi + 
        \epsilon^{1/2}\tilde\eta({\bf k},t).
\end{equation}
The structure factor
$S({\bf k},t)=\langle\tilde\psi({\bf k},t)\tilde\psi(-{\bf k},t)\rangle$
is then,
\begin{equation}\label{EQN:linearS}
        S({\bf k},t) = {\epsilon (e^{2k^2(k_0^2-k^2)t}-1) \over 2(k_0^2-k^2)}.
\end{equation}
$\psi_{\rm rms}$ is found by integrating $S({\bf k},t)$ with
respect to $k$,
\begin{equation}
        \langle\psi^2(t)\rangle=\int {d^dk\over(2\pi)^d} S({\bf k},t).
\end{equation} 
Following \cite{ball}, we observe that the integrand is sharply peaked
about $k=k_0/\sqrt 2$ and we approximate it by a Gaussian.  The
integral is then readily evaluated to find (to leading order in
$1/t$)
\begin{equation}
        \langle\psi^2(t)\rangle= \epsilon\, e^{{1\over 2}k_0^4t}
        \left(\pi\over 2t\right)^{1/2}
        {d\over k_0^{4-d}\Gamma(1+d/2)(8\pi)^{d/2}}.
\end{equation}
This result, when tested against simulations of the fully nonlinear
CHC, agrees well up to $\langle\psi^2(t)\rangle^{1/2}\approx(0.15)$,
thus motivating our choice for $\psi_t$ indicated above.

Finally, we equate $\langle\psi^2(t_0)\rangle$ to $\psi_t^2$ to
determine the time $t_0$ at which the bulk phase separation process
has reached the nonlinear threshold.  The resulting transcendental
equation for $t_0$ can be approximately solved by observing that
$\psi_t^2/\epsilon\gg 1$ in low noise conditions, and that this ratio
must be compensated primarily by the $e^{k_0^4t_0/2}$ factor.
Equating these two and then iteratively improving the estimate of
$t_0$ yields
\begin{equation}\label{EQN:t0}
        t_0 \approx {2\over k_0^4} \ln\left(\psi_t^2\over\epsilon\right)+
        \ln\left(2\Gamma(1+d/2)(8\pi)^{d/2}\sqrt{\ln(\psi_t^2/\epsilon)}\over
                k_0^{d-2}d\right).
\end{equation}

Next we consider the linearized theory for the filled blend.  We solve
(\ref{EQN:CHClin}) for the exterior of the filler particle or fiber,
and in the absence of thermal noise (the noise can simply be averaged
out of the composition wave within the linearized theory).  We
consider idealized filler particles that are symmetric and finite in
some of their coordinates and infinite (i.e., very large on the phase
separation pattern scale) in the remaining coordinates defining the
particle dimensions.  With such symmetry, the composition wave depends
only on the coordinate perpendicular to the interface, which is a radial
coordinate in $d_{\perp}$ dimensions. For example, a spherical particle 
in $d=3$ corresponds to
$d_\perp=3$, a cylindrical fiber is prescribed by $d_\perp=2$, and a
platelet filler reduces to the planar surface with $d_\perp=1$.  We
can treat the general $d_\perp$ case through the
$d_{\perp}$-dimensional Laplacian,
\begin{equation}
  \nabla^2 \psi(r) = {\partial^2\psi\over\partial r^2} + {(d_\perp-1)\over 2}
  {\partial\psi\over\partial r},
\label{laplacian}
\end{equation}
yielding a fourth order partial differential equation for 
(\ref{EQN:CHClin}).  The ``source'' for the composition wave comes from the
boundary conditions obtained by linearization of (\ref{EQN:conservation})
and (\ref{EQN:bcii}), namely
\begin{equation}\label{EQN:conslin}
   \hat r\cdot \nabla(k_0^2 + \nabla^2)\psi(R) = 0
\end{equation}
for conservation at the boundary, and 
\begin{equation}\label{EQN:bciilin}
   \hat r\cdot \nabla \psi(R) = h + g \psi(R).
\end{equation}
This provides a pseudo one-dimensional system which can be readily integrated
numerically.  

\begin{figure}
\hbox to\hsize{\epsfxsize=\hsize\hfil\epsfbox{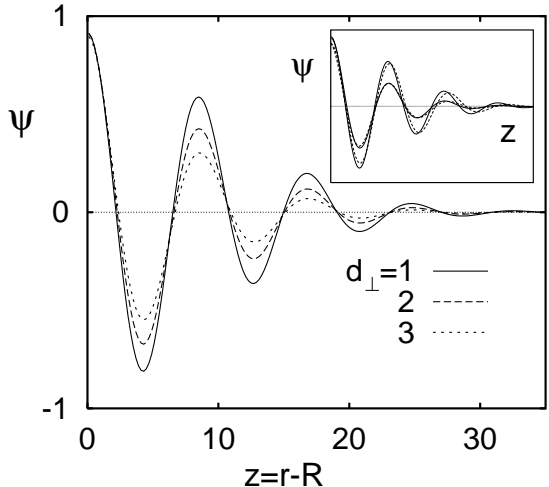}\hfil}
\medskip
 \caption{Influence of particle shape on the development of
 filler-directed composition waves at early time. Radial composition
 profiles $\psi(z)$ are obtained from the linearized CHC equation for
 symmetric particles finite in $d_{\perp}$ directions, corresponding
 to ``platelet'' fillers ($d_\perp=1$), fibers ($d_\perp=2$), and
 spherical fillers ($d_\perp=3$). The radius is $R=10$, and the
 surface interaction parameters are $h=0.005$ and $g=0$. The reduced
 time is $t = 25$.  The inset shows the approximate solutions of
 (\protect\ref{EQN:psiz}) and (\protect\ref{EQN:psiz3}) (dashed lines), 
 compared to
 the $d_\perp=1$ and $3$ numerical solutions from the main figure
 (solid lines).}
\label{FIG:prod}
\end{figure}

The solutions for $d_\perp = 1,2,3$ presented in Fig.~\ref{FIG:prod}
illustrate the influence of $d_{\perp}$ on the composition wave
pattern.  Increasing $d_{\perp}$ reduces the amplitude of the
composition wave. This feature can be understood to arise from the 
increasing volume occupied by the outer rings. The opposite situation
should hold for {\it exterior\/} boundaries having these symmetries,
so that more coherent ring structures might be anticipated in phase
separation confined to these geometries, especially for spherical
cavities. It may prove interesting to examine phase separation in the
presence of fractal filler particles (like fumed silica) to determine
whether geometry stabilizes or destabilizes the phase separation
pattern and how the evolving phase separation pattern accommodates the
fractal boundary structure.

Returning to the analytical estimation of the target pattern size, we
expand $\psi(r)$ in a basis which diagonalizes the Laplacian.  This
amounts to performing a cosine transform for $d_\perp=1$, a ($J_0$)
Hankel transform in $d=2$, and a half-integer Hankel tranform in
$d=3$.  The last can be re-expressed as a Fourier cosine transform of
$r\psi(r)$ rather than $\psi$.  Here we study the extremal cases of
$d_\perp=1$ and $3$, and show that $d_\perp$ has only a minor effect
on the pattern size.

First, we revisit the $d_\perp=1$ case already addressed by Ball and Essery
\cite{ball}.  One can solve (\ref{EQN:CHClin}) with boundary conditions
via Fourier cosine transformation with respect to $r$ and Laplace transform
with respect to $t$, with the result,
\begin{equation}\label{EQN:trans}
  \tilde\psi(k,s) = {hk^2\over s[s-k^2(k_0^2-k^2)]}
\end{equation}
for the case $g=0$ (to which we restrict our attention).  Inverting the
Laplace transform and using a Gaussian approximation again to invert the
Fourier cosine transform gives the solution
\begin{equation}\label{EQN:psiz}
   \psi_{d_{\perp}=1}(z,t) \approx \psi(0,t) e^{-(z/4\lambda_0)^2} 
   \cos (z/\lambda_0), 
\end{equation}
where $\psi(0,t) \equiv {h\over k_0^3}\sqrt{8\over\pi t} \exp\left(
{k_0^4 t\over 4}\right)$, $\lambda_0^{-1} = k_0/\sqrt 2$, and where
subdominant terms in $1/t$ have been neglected.  A non-zero value of
$g$, while complicating (\ref{EQN:trans}), would appear in this
approximate solution only via the substitution $h \to h + g\psi_0$.
In this $d_\perp=1$ example, the $R$ dependence drops out of the
linearized equation, and $z=r$ measures the distance from the wall.

For a given noise strength $\epsilon$ we have a time $t_0$ at which
the local composition of one phase reaches $\psi_t$.  For surface
interaction $h$ we solve for the distance $z_0$ out to which the envelope
of $\psi(r,t_0)$ exceeds $\psi_t$.  This gives the approximation, 
\begin{equation}\label{EQN:z0}
   z_0 \approx 2k_0^3 t_0 \left[1-{1\over k_0^4t_0}\ln\left({k_0^3\psi_t\over
         h}\sqrt{\pi t_0\over 8}\right)\right].
\end{equation}
Thus, the propagation front grows with a velocity $2k_0^3$ at long
times \cite{ball}, and the terms in square brackets are the leading
correction to this long time asymptotic behavior.

The $d_\perp=3$ composition profile may be obtained by observing that
(\ref{linearpsi}) and (\ref{laplacian}) yield the same equation for
$r\psi_{d_\perp=3}(z,t)$ as for the composition profile 
$\psi_{d_{\perp}=1}(z,t)$.  Thus, we impose the boundary conditions 
(to leading order in $R/r$) and follow the above derivation with the result
\begin{equation}\label{EQN:psiz3}
  \psi_{d_\perp=3}(z,t) \approx {R\over R+z} \ \psi_{d_\perp=1}(z,t).
\label{fit}
\end{equation}
Solving for the value $z_0$ at which the
composition wave envelope equals $\psi_t$ leads to nearly the same
expression as the $d_\perp=1$ case (\ref{EQN:z0}), with an
additional $-\ln(1+z_0/R)/k_0^4t_0$ term in the square brackets.
Typically $z_0\lesssim 10R$, $k_0$ is of order unity, and $t_0$ ranges
from 10 to 30, making this term roughly a 10\% correction.

To compare with our simulations, we consider $d_\perp=d=2$.  For $z_0$
we simply take the arithmetic mean of the values obtained for
$d_\perp=1$ and $d_\perp=3$ (motivated by numerical solutions). 
Equation (\ref{EQN:t0}) for
$t_0$ is substituted into (\ref{EQN:z0}) to obtain a prediction for
$z_0$ in terms of $h$, $\epsilon$, and $\psi_0$.  Here we consider
critical quenches with $\psi_0=0$, or $k_0=1$.  Finally, we
approximate $\ln t_0$ and $\ln z_0$ with typical values, which
introduces less than 10\% error with the range of parameters
considered here, and thus obtain
\begin{equation}\label{EQN:pred}
  z_0 \approx -2.5 - 8.5 \log_{10}\epsilon + 4.6 \log_{10} h.
\end{equation}
A similar expression results for $d=d_\perp=3$, with the primary difference
being a change of the $\log_{10}\epsilon$ coefficient to $-7.9$.

\begin{figure}
\hbox to\hsize{\epsfxsize=1.0\hsize\hfil\epsfbox{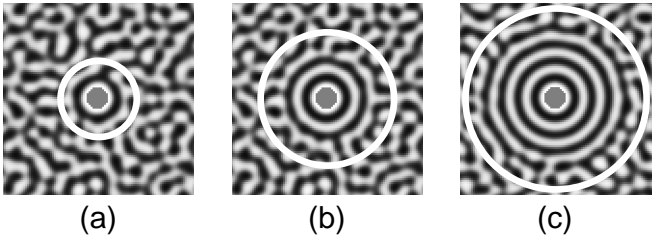}\hfil}
\vspace{5mm}
\caption{Comparison between analytic estimate of the spatial extent of
the target pattern the corresponding CHC simulation at an intermediate 
stage of phase
separation. Simulations are performed at noise levels (a)
$\epsilon=10^{-2}$ corresponding to modest molecular weight and
shallow temperature quenches, (b) $\epsilon=10^{-4}$, and (c)
$\epsilon= 10^{-6}$, corresponding to high molecular weight mixtures
and deep temperature quenches. The surface interaction $h=1$ for all
cases, the radius $R=5.6$, and the times for each quench are chosen to
correspond to when the spinodal pattern is fully developed: (a)
$t=20$, (b) $t=30$, and (c) $t=40$.  The solid lines are the
predictions of (\protect\ref{EQN:pred}).}
\label{FIG:estimator}
\end{figure}

Fig.~\ref{FIG:estimator} shows a comparison of this estimate with the
simulations. We see that the analytic approximation provides a good
rough estimate of the spatial extent of the phase separation pattern,
although it predicts a size typically one oscillation larger than the
outermost unbroken target.

\begin{figure}
\hbox to\hsize{\epsfxsize=1.0\hsize\hfil\epsfbox{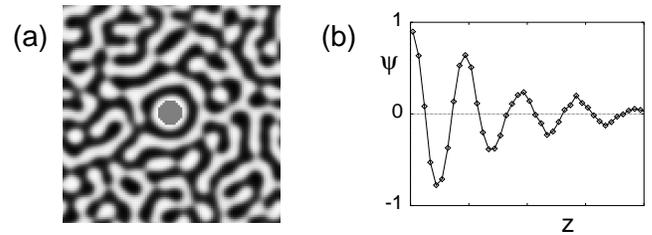}\hfil}
\medskip
\caption{ Late stage target pattern and corresponding radial
composition profile.  a) Filler-induced phase separation pattern for
an intermediate noise value of $\epsilon=10^{-3}$ at time $t=35$,
$h=1$, and $R=5.6$.  b) Averaged radial profile about center of filler
particle. Note that the ring composition pattern persists in the
radial average even after the target pattern appears visually to have
broken up.}
\label{FIG:averaged}
\end{figure}

We point out that the spherical composition waves are apparent in the
average composition profiles even in the rather noisy looking ring
patterns found in the late stage of target pattern formation.  In
Fig.~\ref{FIG:averaged} we show a target pattern at intermediate
values of noise, as well as the radial average of the composition
profile about the center of the target.  Comparison shows that the
ring composition pattern persists in the radial average even after the
target pattern appears visually to have broken up.  This provides a
possible explanation of the apparent overestimate of the target size
in Fig.~7.

\section{Comparison with Experimental Results}

While scattering measurements of the growth of composition waves are
readily performed for a single plane boundary \cite{krausch}, these
measurements become more difficult in filled blends where the filler
particles are randomly distributed within the blend.  This situation
is unfortunate, given the predicted transient nature of the
composition wave patterns when the particles are small. However, real
space studies of blend phase separation are possible in films
sufficiently thin to suppress the formation of surface-directed waves
normal to the solid substrate \cite{sung}.  Under the favorable
circumstances that one of the polymer components segregates to both
the solid substrate and the polymer-air boundary of these nearly
two-dimensional (``ultrathin'') blend films, phase separation is
observed within the plane of the film \cite{karim,sung,ermi}. The
variation of the surface tension in the film accompanying phase
separation gives rise to film boundary undulations that can be
measured by atomic force microscopy (AFM) and optical microscopy (OM)
\cite{sung,ermi}.  The thickness of ultrathin blend films is typically
restricted to small values ($L \approx 200$ nm) and the height
contrast of the surface patterns tends to become larger in still
thinner films \cite{nisato}. A film thickness in the range of 20 -- 50
nm is often suited for observing well resolved phase separation
surface patterns similar to those found in simulations of bulk
blends. In the following we compare our results with those of a model
blend utilized in ultrathin phase separation studies reported
elsewhere with silica beads added as the model filler \cite{karim2}.

The spun-cast films are composed of a near-critical composition blend
of polystyrene and poly(vinyl methyl) ether (PVME). The filler
particles are silica beads having an average size of about 100 nm, as
measured by direct imaging of the particles.  This particular filler
was chosen because of its tendency to be enriched by polystyrene,
rather than PVME, which enriches both the solid and air surfaces. In
this way, the filler particles are not competing with the solid or air
surfaces for the enriching polymer.  Phase separation was achieved by
annealing the film approximately 15$^\circ$ within the two-phase
region, corresponding to a fairly shallow quench.  Film topography
(height) was measure by AFM.  Further details of the experiment are
provided in Ref.~\cite{karim2}.

\begin{figure}
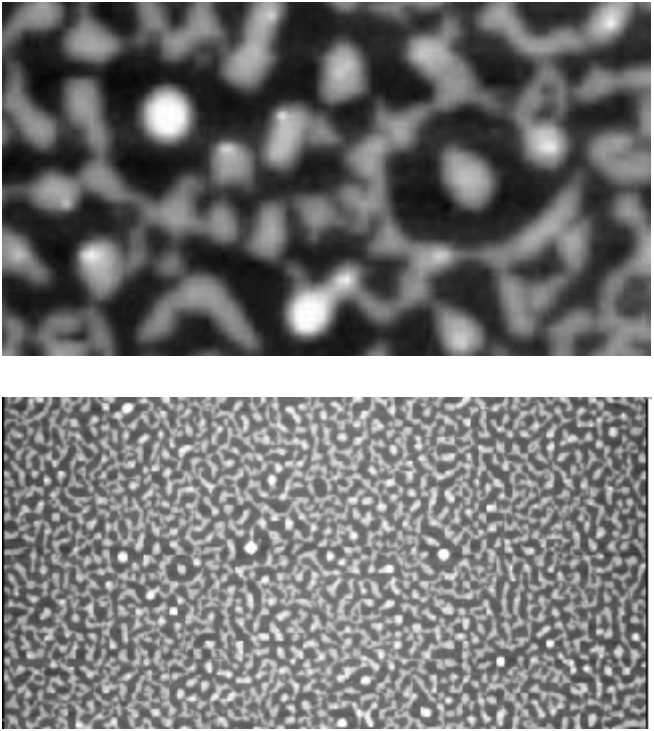

\hbox to\hsize{\epsfxsize=1.0\hsize\hfil\epsfbox{ZoomTrgt-half.epsi}\hfil}
\vspace{5mm}
\hbox to\hsize{\epsfxsize=1.0\hsize\hfil\epsfbox{Targ4-bw-half.epsi}\hfil}
\vspace{5mm}
\caption{AFM image of phase separation pattern in PS/PVME blend film
with dilute concentration of filler particles at (top) 20$\mu$m and
(bottom) 100 $\mu$m scale. The height undulations reflect composition
variations within the film associated with surface tension variations
\protect\cite{ermi}. Image contrast has been enhanced by a film washing
procedure.  From Ref.~\protect\cite{karim2}.}
\label{figkarim}
\end{figure}

Fig.~\ref{figkarim}(top) shows the topography of the blend film at an
intermediate stage of phase separation where we expect circular
filler-induced composition waves to be evident. The pattern resembles
the simulated patterns under similar quench conditions.  The symmetry
of the film phase separation pattern is locally broken by the presence
of the filler particles, leading to the formation of ring-like
concentration wave patterns. Note that when observed on a larger scale
[Fig.~\ref{figkarim}(bottom)], the phase separation pattern far from
any filler particles resembles the typical spinodal decomposition
pattern observed in control measurements on the same blends without
filler.

\begin{figure}
\hbox to\hsize{\epsfxsize=0.9\hsize\hfil\epsfbox{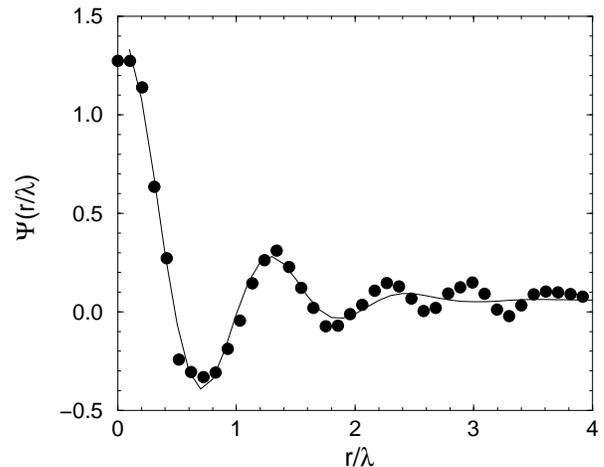}\hfil}
\vspace{5mm}
\caption{Radial average of a two-dimensional fast Fourier transform of
AFM height data centered about an isolated target pattern in 
Fig.~\protect\ref{figkarim}.
The scale of the damped oscillatory profile is reduced by the phase
separation scale determined by a Fourier analysis of the AFM height
data for the entire film.  The solid line is a fit to the data as
described above.  Data from Ref.~\protect\cite{karim2}.}
\label{figradial}
\end{figure}

It is apparent that the patterns in Fig.~\ref{figkarim} are in a
relatively late stage of phase separation, where the rings are
beginning to break up along with the ``background'' phase separation
pattern.  The simulations above indicate that the target patterns are
more persistently expressed in the radially averaged patterns and in
Fig.~\ref{figradial} we show the radial average of the AFM height data
centered about a representative filler particle. The target pattern in
the radially averaged data extends far beyond the ring feature
apparent in the image in Fig.~\ref{figkarim}.  The data in
Fig.~\ref{figradial} correspond to a shallow quench, and are
comparable to the intermediate stage, shallow quench simulation data
in Fig.~\ref{FIG:proR}.

Next we directly compare the prediction of the linearized theory to
the AFM data.  An exact solution of $\psi_{d_{\perp}=2}(z,t)$ is
difficult, but we can obtain a reasonable approximation to
$\psi_{d_{\perp}=2}(z,t)$ by generalizing the method described above
for $\psi_{d_{\perp}=1}(z,t)$.  We estimate $\psi_{d_{\perp}=2}(z,t)$
as a Gaussian decay function multiplied by the eigenfunction of the
Laplacian in $d=2$ [rather than in $d=1$ as in the case of
(\ref{EQN:psiz})]. In this approximation, $\psi_{d_{\perp}=2}(z,t)$
becomes a product of a Gaussian as in Eq.~(\ref{EQN:psiz}) and a Bessel
function $J_0(2\pi z/\lambda_0)$, and we show a fit of this function
to the AFM data in Fig.~\ref{figradial}. The fitted value of the
particle radius $R$ is $82$ nm, which is comparable to the average
particle radius obtained by optical microscopy ($R \approx 100$ nm).
The scale parameters of the Gaussian and Bessel functions have been
adjusted along with the prefactor which is set by the value of
$\psi_{d_{\perp}=2}(z,t)$ as $z$ tends to zero. It is clear that the
oscillatory pattern scale is on the order of the background phase
separation pattern, and that the linearized expression for
$\psi_{d_{\perp}=2}(z,t)$ has the qualitative shape of the measured
profile. Such qualitative agreement is the best that can be expected
from the linearized theory, which strictly speaking should hold only
at very early times.

At still longer times, the phase separation pattern eventually breaks up
into droplets and little difference is observed between the films with
and without filler. Thus, the target patterns induced by the filler
particles are transient, as observed in the simulations. Of course,
the version of the CHC model used here cannot reliably describe
quantitative features of these late stage processes without the
incorporation of hydrodynamic interactions.

Under far off-critical conditions and a selective interaction between
the filler particles and one of the polymers ($h>0$) the filler
particles are ``encapsulated'' by a layer of the favored polymer so
that concentration waves do not develop. The formation of droplets by
nucleation or far off-critical spinodal decomposition can also have
the effect of breaking the symmetry of the phase separation process,
but the pattern formation is not generally the same as for critical
composition mixtures. Recent measurements have reported the occurrence
of filler encapsulation in a blend of polypropylene and polyamine-6
with glass bead filler particles \cite{qui1,benderley}. Encapsulation
occurs when the polypropylene-rich phase having the selective
interaction for the filler is the minority phase, but no encapsulation
occurs when polypropylene is the majority phase. This finding compares
well with the simulation results discussed in section IV.

\begin{figure}
\hbox to\hsize{\epsfxsize=0.9\hsize\hfil\epsfbox{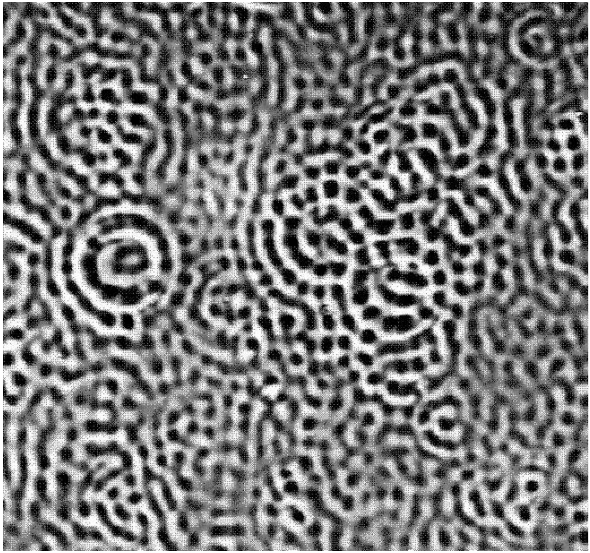}\hfil}
\vspace{5mm}
\caption{Phase contrast microscopy image of phase separation pattern of a
photo-crosslinked blend. The scale of the image is 10 $\mu$m. From
\protect\cite{qui1}.  Reproduced with permission from Marcel Dekker.}
\label{fig12}
\end{figure}

Radiation crosslinking provides another source of heterogeneity that
can be introduced readily in phase separating films.  Measurements of
irradiated photoreactive blends of PVME and PS with a crosslinkable
side group styrene-chloromethyl styrene random copolymer (PSCMS) show
the formation of striking ring composition patterns \cite{tamai,qui2}
and we reproduce one of these patterns in Fig.~\ref{fig12} (compare
with Fig.~\ref{FIG:multiple}).  Furukawa \cite{furukawa} has
interpreted these observations in terms of a model by which
irradiation first brings the blend into the nucleation regime where
droplets phase separate, followed by the entrance into the spinodal
regime where the droplets act like the filler particles discussed in
the present paper. This is a plausible interpretation of the
qualitative origin of these patterns, but it is difficult to interpret
these measurements directly from CHC simulations since crosslinking
imparts a non-trivial viscoelasticity to the polymer blend
\cite{onuki}. The crosslinking, which also increases the molecular
weight of PSCMS, and the increased elasticity, both lead us to expect
a decrease in the thermal noise and thus an increased tendency to form
target patterns.  It also seems plausible that the crosslinks
themselves provide the source of heterogeneity, inducing the
development of composition waves.
              
\section{Discussion} 

The presence of filler particles in a phase separating fluid mixture
is found to give rise to transient composition wave patterns in
simulations based on the Cahn-Hilliard-Cook model and in measurements
\cite{karim2} on ultrathin polystyrene/poly(vinylmethyl) ether blend
films with silica filler particles. In both the simulation and the
experiment, the composition wave patterns were found to be transient
and the filler is found to have a diminishing effect as the scale of
the phase separation pattern becomes larger than the filler
particles. The propagation of composition waves is enhanced at lower
thermal noise level so that the effect propagates to larger distances
for deeper quenches and higher molecular weight blends. The finite
size and the dimensionality of the filler particles are found in our
simulations to have a similar effect in determining the stability of
the composition wave pattern at intermediate times. The composition
waves become more stable for particles large in comparison to the
spinodal wavelength and the concentration waves exhibited by these
larger particles are similar to planar interfaces. The composition
waves about the filler particles are more stable for particles
extended at great distance along more directions; i.e., surfaces are
more stable than long cylinders, which are more stable than spherical
filler particles. Our results compare favorably with experiments on
phase-separating filled blend films which are nearly two-dimensional.

Filler particles are an example of a perturbation of phase separation
by boundaries {\it interior} to the fluid. It would be interesting to
investigate the influence of exterior boundaries of finite extent on
phase separation. It seems likely that composition waves within
confined geometries should be more stable because of the decreasing
surface area of the rings farther from the surface. This should lead
to well developed and more long lasting perturbations of the phase
separation process. The relation between boundary shape and phase
separation morphology should be very interesting for this class of
measurements.  Phase separation within arrays of filler particles
where the distinction between interior and exterior boundaries becomes
blurred and where larger perturbations of the phase separation process
may be anticipated, should also prove interesting. The distinction
between large and small and fixed and mobile filler particles should
lead to a range of new phase separation morphologies since the
development of composition waves should lead to changes in the
filler-filler interaction that can influence the subsequent
development of the film structure. The utilization of geometrically
and chemically patterned surfaces and additives offer many
opportunities for the control of the phase separation morphology and
resulting properties of blend films, and the study of these
surface-induced phase separation processes raise many interesting
problems of fundamental and practical interest.

\acknowledgments

The present work has benefitted greatly from close collaboration with
experimentalists in the Polymer Blends Group at NIST. The simulations
and experimental work were conducted simultaneously, and we thank
Alamgir Karim and Eric Amis for many suggestions which influenced the
design and interpretation of our simulations.  We thank Giovanni
Nisato for many useful conversations and providing the correlation
function data for the filled films.  We have also benefitted from
conversations with Qui Tran-Cong regarding the relation of his
measurements to our simulations and for contributing Fig.~11.
B.P. Lee acknowledges the support of the National Research
Council/NIST postdoctoral research program.

\appendix

\section{Boundary Conditions on a Curved Surface}  

Curved boundaries complicate the implementation of boundary conditions
in a spatially discretized simulation.  In the present work, we use a
square lattice and simulate filler particles with circular,
cylindrical, and spherical symmetry.  This requires a method of
incorporating the boundary conditions that minimizes the effects of
errors caused by approximating curved boundaries by lattices.  In this
appendix we present our approach to this problem.

\begin{figure}
\hbox to\hsize{\epsfxsize=0.7\hsize\hfil\epsfbox{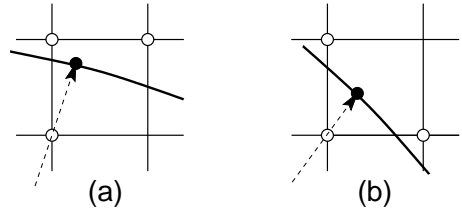}\hfil}
\vspace{5mm}
\caption{A curved boundary passes through either adjacent or opposite
sides of a lattice unit cell.  For both cases we express the field 
and its normal derivative at $(x_0,y_0)$, the solid dot, in terms of the
values at the three vertices depicted by open circles.  The dashed line
is the radius of the boundary arc.}
\label{FIG:boundary}
\end{figure}

Generally, (\ref{EQN:conservation}) and (\ref{EQN:bcii}) are imposed
by inclusion of $\psi$ and $\mu=-\nabla^2\psi-\psi+\psi^3$ values at
the lattice sites on the immediate interior of the boundary (within
the wall or filler), which are determined from the boundary conditions
before each time step.  We superimpose the circular boundary over the
square lattice so that no lattice vertices lie along the boundary.
Consequently, every interior point corresponds to one of two
possibilities, shown as the lower left corners of
Fig.~\ref{FIG:boundary} (a) and (b).  The boundary condition at the
point $(x_0,y_0)$ (shown as a black dot) is not set at the interior
lattice site but rather at the intersection of the boundary and the
radius passing through the interior lattice site.
                
In both cases of Fig.~\ref{FIG:boundary} we use the three vertices 
shown as open circles for the discretized representation of $\psi(x_0,y_0)$
and its normal derivative $\hat r\cdot\nabla\psi(x_0,y_0)$.  To highest
order these representations are unique [to $O(\Delta x^2)$ for $\psi$ and  
$O(\Delta x)$ for the derivative, with $\Delta x$ the lattice spacing]. 
Hence, (\ref{EQN:bcii}) may be used to determine $\psi_{i,j}$, the field at
the interior site, from the appropriate exterior points.  We find for
case (a) the relation
\begin{equation}
        \psi_{i,j}  = { (1-g\ell)\bigl[ (\sin\theta - \cos\theta)\, \psi_{i,j+1}
                           + \cos\theta\, \psi_{i+1,j+1}\bigr] -h\Delta x \over
                        (1-g\ell)\sin\theta + g\Delta x },
\end{equation}
where $\ell$ is the distance from the 
interior lattice site to $(x_0,y_0)$ while $\theta$ is the angle between the
radius and the horizontal axis.  For case (b) the analogous expression is
\begin{equation}
        \psi_{i,j}  = { (1-g\ell)\bigl[\cos\theta\, \psi_{i+1,j}
                        +\sin\theta \, \psi_{i,j+1}\bigr] - h \Delta x \over
                        (1-g\ell)(\cos\theta + \sin\theta) + g\Delta x }.
\end{equation}

The chemical potential $\mu_{i,j}$ may also be assigned at the interior 
point, in practice by assigning a value to $(\nabla^2\psi)_{i,j}$ to supplement
the Laplacian derived from $\psi$ outside the boundary.
In this way we can impose the conservation requirement 
(\ref{EQN:conservation}) for case (a) via 
\begin{equation}
        \mu_{i,j} = (1-\cot\theta)\, \mu_{i,j+1} + \cot\theta\, \mu_{i+1,j+1},
\end{equation}
while for case (b),
\begin{equation}
        \mu_{i,j} = {\cos\theta\over\cos\theta + \sin\theta} \,\mu_{i+1,j} +
                   {\sin\theta\over\cos\theta + \sin\theta} \,\mu_{i,j+1}.
\end{equation}
In simulations with thermal noise we assume a separation of time scales
between thermal fluctuations and order parameter variations (as described
in \cite{marko}), and simply supplement the above conditions with the
conservation law for fluctuations at the boundary: $\hat r \cdot \nu =0$,
where $\nu$ is the noise current derived from 
$\eta= \nabla \cdot \nu$.


\begin{references}
\bibitem{chempot} The chemical potential $\mu_A - \mu_B = \mu_{eq}$ is
often omitted in (\ref{EQN:functional}), but is generically
non-zero, except in the special case of $N_A=N_B$ and $\phi = \phi_c$.
\end{references}
\end{document}